# Tip cell overtaking occurs as a side effect of sprouting in computational models of angiogenesis

Sonja E. M. Boas[1,2] and Roeland M.H. Merks[1,2]

1. Life Sciences, Centrum Wiskunde & Informatica (CWI), Science Park 123, 1098XG Amsterdam, The Netherlands;
2. Mathematical Institute, Leiden University, Niels Bohrweg 1, 2333 CA Leiden, The Netherlands

Email addresses: Roeland.merks@cwi.nl and boas@cwi.nl

Corresponding author: Roeland.merks@cwi.nl

## Abstract

***Background*** During angiogenesis, the formation of new blood vessels from existing ones, endothelial cells differentiate into tip and stalk cells, after which one tip cell leads the sprout. More recently, this picture has changed. It has become clear that endothelial cells compete for the tip position during angiogenesis: a phenomenon named tip cell overtaking. The biological function of tip cell overtaking is not yet known. From experimental observations, it is unclear to what extent tip cell overtaking is a side effect of sprouting or to what extent it is regulated through a VEGF-Dll4-Notch signaling network and thus might have a biological function. To address this question, we studied tip cell overtaking in computational models of angiogenic sprouting in absence and in presence of VEGF-Dll4-Notch signaling.

***Results*** We looked for tip cell overtaking in two existing Cellular Potts models of angiogenesis. In these simulation models angiogenic sprouting-like behavior emerges from a small set of plausible cell behaviors. In the first model, cells aggregate through contact-inhibited chemotaxis. In the second model the endothelial cells assume an elongated shape and aggregate through (non-inhibited) chemotaxis. In both these sprouting models the endothelial cells spontaneously migrate forwards and backwards within sprouts, suggesting that tip cell overtaking might occur as a side effect of sprouting. In accordance with other experimental observations, in our simulations the cells' tendency to occupy the tip position can be regulated when two cell lines with different levels of *Vegfr2* expression are contributing to sprouting (mosaic sprouting assay), where cell behavior is regulated by a simple VEGF-Dll4-Notch signaling network.

***Conclusions*** Our modeling results suggest that tip cell overtaking can occur spontaneously due to the stochastic motion of cells during sprouting. Thus, tip cell overtaking and sprouting dynamics may be interdependent and should be studied and interpreted in combination. VEGF-Dll4-Notch can regulate the ability of cells to occupy the tip cell position in our simulations. We propose that the function of VEGF-Dll4-Notch signaling might not be to regulate which cell ends up at the tip, but to assure that the cell that randomly ends up at the tip position acquires the tip cell phenotype.

## Key words

# 1. Background

Oxygen deprived regions, such as wounds and growing tumors, can stimulate the sprouting of side branches from nearby vessels, a process called angiogenesis [1]. Growth factors activate quiescent endothelial cells, which differentiate into one of two alternative fates: a 'tip cell' or a 'stalk cell' [2-4]. Tip cells are the initiators and leaders of new sprouts, while stalk cells form the body of the new sprout. Activated endothelial cells compete for the tip cell fate through lateral inhibition by Dll4-Notch signaling, a process called tip cell selection [2-4]. In this process, tip cells present Dll4 ligands on their membrane to activate Notch receptors of their neighbors. Upon Notch activation, the Notch-intracellular domain (NICD) is cleaved off and travels to the nucleus for transcription of Notch target genes. Eventually, cells with low Notch activity (low Notch/high Dll4) become tip cells and cells with high Notch activity (high Notch/low Dll4) become stalk cells. Previous work assumed that the tip cell at the sprout front maintained its leader position during sprouting [3]. More recently, Jakobsson *et al.* [5] and Arima *et al.* [6] showed independently that cells compete for the tip position of sprouts during angiogenesis, a phenomenon named tip cell overtaking. The biological relevance of tip cell overtaking is not yet clear. In this paper we use computational modeling to study if tip cell overtaking is merely a side effect of sprouting, or if it is regulated by intercellular signaling and thus likely has a regulatory function in sprouting.

Jakobsson *et al.* [5] and Arima *et al.* [6] both observed tip cell overtaking in sprouting assays, but they interpreted their data differently with respect to the regulation of tip cell overtaking. Using genetic mosaic sprouting assays, Jakobsson *et al.* [5] found that cells with relatively high levels of *Vegfr2* expression or relatively low levels of *Vegfr1* expression are more likely to end up at the tip position in a Notch-dependent fashion, suggesting that the competitive potential of cells to take up the tip position is regulated by the signaling networks consisting of VEGF, Dll4 and Notch. VEGF influences tip cell selection by inducing Dll4 production upon VEGFR2 activation [7]. Notch activation in neighboring cells down-regulates *Vegfr2* expression [8]. Using this signaling network, computational modeling by Jakobsson *et al.* [5] suggested that tip cell overtaking is regulated by Notch activity. In a follow-up model, Bentley *et al.* [9] studied the role of cell-cell adhesion and junctional reshuffling, using a variant of the Cellular Potts Model, allowing cells to crawl along one another within a preformed cylindrical hollow sprout. By comparing different combinations of mechanisms, their modeling results suggested a more detailed regulatory mechanism for tip cell overtaking: 1) VEGFR2 signaling causes endocytosis of VE-cadherin, which reduces cell-cell adhesion. 2) Notch activity decreases extension of polarized actomyosin protrusions towards the sprout tip. Thus, these results suggest that Dll4-Notch and VEGF signaling strongly regulate tip cell overtaking.

In apparent contradiction with this interpretation, Arima *et al.* [6] found that tip cell overtake rates were not affected by addition of VEGF or by inhibition of Dll4-Notch signaling, although other measures of sprouting kinetics were influenced, e.g., sprout extension rate and cell velocity. Arima *et al.* [6] presented extensive cell tracking data of cell movement and position during angiogenic sprouting and found that individual ECs migrate forwards and backwards within the sprout at different velocities, leading to cell mixing and overtaking of the tip position. Thus, tip cell overtaking might arise spontaneously from collective cell behavior driving angiogenic sprouting.

To help interpret these results, we first studied to what extent tip cell overtaking occurs in existing computational models, without making any additional assumptions (Figure 1A). Although the exact cellular mechanisms driving angiogenesis are still incompletely understood, a range of computational models has been proposed each representing an alternative, often related mechanism [10, 11]. In absence of a definitive sprouting model, we compared two previous Cellular Potts models [12, 13]. In the first model, the cells secrete a chemical signal that attracts surrounding cells *via* chemotaxis. Portions of the membrane in contact with adjacent cells become insensitive to the chemoattractant [13]. The model forms sprouts of one or two cell diameters thickness (Figure 2A and 2C). The second model hypothesizes that non-inhibited chemotaxis suffices to form angiogenesis-like sprouts, if the cells have an elongated shape [12] (Figure 2B and 2D).

      As a second step, we studied how Dll4-Notch and *Vegfr2* expression can bias cells to the tip position in these sprouting models (Figure 1B). We introduced a modified existing model of the VEGF-Dll4-Notch signaling network [14] into each simulated cell, and asked to what extent such molecular signaling can fine-regulate tip cell overtaking.

## 2. Results
### 2.1 Spontaneous tip cell overtaking in computational models of angiogenic sprouting

To study if tip cell overtaking can arise spontaneously as a side effect of sprouting, we used two computational models in which sprouts form autonomously, in absence of any type of tip cell selection or regulation. We will briefly introduce both models here, referring to the Method Section 5.1 and previous publications [12, 13] for details. Both models consider a restricted set of cell behaviors to explain the autonomous growth of angiogenic sprouts from an initial spheroid of endothelial cells. Both models assume that endothelial cells attract one another via a secreted, diffusive, short-lived chemical signal, forming exponential chemoattraction gradients, e.g., via isoforms of VEGF diffusing over one to few cell diameters. This assumption produces aggregates of endothelial cells [12, 13, 15], but it will form networks of cells with an additional assumption. The 'contact inhibition model' [13] (Figure 2A), additionally proposes that chemotaxis is inhibited at cell-cell interfaces, i.e., they only chemotact at cell-extracellular matrix interfaces. The effect might be due to VE-cadherin-signaling, with VE-cadherins interacting locally with VEGFR2 [16]. The 'cell elongation model' [12] (Figure 2B) showed that the elongated shape of endothelial cells suffices for network formation. In variants of this model cells attract one another via weak cell-cell adhesion [17] or via a longer range potential [18].

In order for VEGF to serve as an attraction signal, its diffusion coefficient must be sufficiently low or the degradation rate sufficiently high so it can form gradients with a diffusion length of one to a couple of cell diameters. This contradicts with VEGF's role as a long-range cue guiding blood vessels over longer distances; e.g., hypoxic tumors can attract over distances up to 2-3 mm [19]. A recent model [20, 21] and experimental observations [21] suggest that secreted VEGF accumulates close to the endothelial cells and colocalizes with fibronectin and heparin sulfate proteoglycan. Thus although the diffusion length of soluble VEGF is longer than what was assumed in these computational models, binding to the extracellular matrix may strongly reduce the diffusion rate of VEGF and create much shorter gradients of ECM-bound VEGF near the endothelial cells. This role of VEGF as a short-range attractive signal differs from the role of VEGF as a long range guidance cue. For the purpose of this paper, chemo-attraction is considered representative for other potential attraction mechanisms including cell-cell adhesion [22, 17] or mechanotransduction via the extracellular matrix [23, 24]. The insights do not depend on the precise mechanism of the attractive forces between endothelial cells.

Spontaneous tip cell overtaking occurs in both models as a side effect of sprouting. Figure 2C shows an example of tip cell overtaking in the contact inhibition model. The cell labeled with a green dot overtakes the cell labeled with a gray dot. Figure 2D shows an example of a tip cell overtake in the cell elongation model, where the cell labeled with a purple dot overtakes the cell labeled with a pink dot. In our recent model of mechanical cell-cell communication via the extracellular matrix [23], tip cell overtaking rarely occurred; we therefore did not study tip cell overtaking in this model.

### 2.2 Quantification of tip cell overtaking

To quantify tip cell overtaking during sprouting in the contact inhibition model and in the cell elongation model, we first identified the cell on the sprout tip, 'the leader cell'. The leader cell of each sprout is identified at each time step (Monte Carlo Step, MCS) of the simulations, using an automated method (see Section 5.2). Figure 2A and 2B show a vascular network formed by the contact inhibition model and the cell elongation model with the leader cells colored in red. Tracking of the leader cells allowed us to identify overtaking events. We define a tip cell overtake as the replacement of a leader cell by a neighboring cell. To prevent overestimates of tip cell overtake events due to the short-lived, random cell protrusions that the Cellular Potts describes, an overtake is counted only if both the leader cell and the overtaking neighboring cell have been present at the tip position for at least 80 consecutive MCS. Assuming that 1 MCS corresponds to thirty seconds, we thus count overtake events lasting for longer than forty minutes.

To quantify the frequency of tip cell overtaking, the mean overtake rate per sprout of a simulation was calculated by dividing the number of overtakes within each sprout by the total number of sprouts present in the simulation between MCS 10000 and 30000; i.e. over a period of 7 days with the assumed time scaling of 1 MCS = 30 s. The calculation started from MCS 10000, since sprouts are then well formed from the initial spheroid and the overtake rate was averaged over fifteen independent simulations with the reference parameter settings. Within the time period of 7 days we identified on average $0.67 \pm 1.32$ overtake events in the contact inhibition model. Within the same simulated interval, we identified on average $4.59 \pm 5.24$ overtakes in the cell elongation model. Thus, the average tip cell overtake rate for the cell elongation model is significantly higher than for the contact inhibition model ($p=0.0089$ using an unpaired t-test). There are two explanations for the higher tip cell overtake rate in in the cell elongation model compared to the contact inhibition model. First, in the cell elongation model, aligned elongated cells in a multi-cellular sprout tip can easily slide past another to overtake the tip position, whereas in the contact-inhibition model cells must pass one another completely to establish a tip cell overtake. Second, sprouts in the cell elongation model have longer life-times. In the contact inhibition model sprouts often fuse by anastomosis, resulting in sprouts with short life-times and often lacking a tip cell overtake.

In addition to the tip cell overtake frequency per sprout, we measured the average life-time of tip cells in sprouts for both models. In the contact inhibition model tip cells persist on average for $442 \pm 361$ minutes and in the cell elongation model on average for $1372 \pm 1417$ minutes. Interestingly, the cell elongation model has a higher tip cell overtake frequency in combination with a higher tip cell duration compared to the contact inhibition model. This can be explained by the shorter life-time of sprouts in the contact inhibition model due to frequent anastomosis, thereby often producing short-lived sprouts (and tip cells) not associated with tip cell overtake events. The tip cell overtake rates found in our models of around one per 7 hours to one per 23 hours are of the same order as those observed in experiments [6, 5]. Arima *et al.* [6] measured an interval of approximately 6 to 15 hours for the overtaking of tip cells and Jakobsson *et al.* [5] measured an interval of 3.7 hours, but note that this similarity between model and experiment depends on our choice of the time scaling of the cellular Potts model (CPM).

The mean tip cell overtake rate in both models is robust to changes in parameter values of most of the main parameters of the models (Figure S1 and S2). In the contact inhibition model however, the tip cell overtake rate is sensitive to the level of cell-cell adhesion. In summary, these results show that tip cell overtake events can occur in both models based on intrinsic cell behaviors as a side effect of sprouting, in absence of Dll4-Notch signaling or other molecular regulation.

## 2.3 Simulations suggest that sprouting drives cell mixing and tip cell overtaking
Jakobsson *et al.* [5] and Arima *et al.* [6] have both tracked cell movement during sprouting and showed that individual cells migrate forwards and backwards in sprouts, resulting in shuffling of cells within the sprout, called cell mixing. In this light, tip cell overtaking could be seen as cell mixing specifically at the tip of the sprout. We therefore studied if cell mixing occurred spontaneously in the sprouts formed in the contact inhibition model and in the cell elongation model. Figure 2C and 2D already showed that cell mixing occurs in both models, as the leader cells in the first time frame are both overtaken and subsequently migrate backwards in the sprout. Supplementary movies S1 and S2 show tip cell overtakes in time for the contact inhibition model and for the elongation model, respectively. To study cell mixing in more detail, Arima *et al.* [6] used time-lapse microscopy to track the position of each cell in a sprout over time and quantified their movements. They proposed a range of measures, including: *coordination* (angle between the direction of cell movement and the direction of sprout elongation) and *directional motility* (percentage of cells moving anterograde or retrograde).

We performed an identical analysis for the contact inhibition model and the cell elongation model. A sprout is defined as the leading cell together with its ten nearest neighbors in the same sprout (see Methods Section 5.3). Figure 3A-C show the position of cells relative to the axis of elongation (see Methods Section 5.3) of a sprout in time, for an experiment by Arima *et al.* [6] (Figure 3A), for the contact inhibition model (Figure 3B) and for the cell elongation model (Figure 3C). The cell with the highest positional index represents the tip cell. Overtakes of tip cells can be seen in Figures 3A-C, as each figure contains at least one intersection of a line representing the position of a competing cell with the line that represents the position of the overtaken tip cell. Additionally, each figure shows cells migrating forwards and backwards (cell mixing) within the sprout. For example, the leader cell in the contact inhibition model at 400 minutes of sprouting time migrates backwards in the sprout as indicated by the decrease in position of this cell in Figure 3B, with five cells in front of it at 1600 minutes.

Forward and backward movement is expressed by *coordination*, defined as the average angle ($\theta$) of cell movement with the sprout elongation axis measured each 20 MCS. Figure 3D shows the standard deviation of the pooled time series of $\theta/\pi$ for anterograde moving cells and Figure 3E for retrograde moving cells, showing similar values for experimental and computational results. Similar to *directional motility* in the experimental observations, the majority of the cells is moving forwards ($\theta>\pi$) or backwards ($\theta<\pi$) in both models (Figure 3F). Only a small portion of the cells is not moving, this 'stopped' cell fraction is smaller in the models than in the experiments, indicating that cells in the model are a bit more motile than in the experiments.

Inspired by the notion of cell mixing, we asked whether cell movement during sprouting follows a random walk along the sprouting axis. For this purpose, the centers of mass of the cells were tracked during sprouting and projected on the sprout elongation axis (see Methods Section 5.3). Figure 3G and Figure 3H show the one-dimensional mean square displacement of cells during sprouting in the contact inhibition model and in the cell elongation model, respectively. From the MSD over sprouting time, one can derive that cells move by a biased random walk during sprouting, with a dispersion coefficient of $0.0021 \pm 1.2\cdot10^{-5}$ $\mu m^2/s$ in the contact inhibition model and of $0.0086 \pm 5.1\cdot10^{-5}$ $\mu m^2/s$ in the cell elongation model (see Methods Section 5.3). The dispersion coefficient for cells in the cell elongation model is slightly overestimated since small protrusions by an elongated cell can cause a large position change for its center of mass.

In summary, these results show that all cells in the sprouts behave as random walkers, moving forwards and backwards along the sprout, resulting in cell mixing. Cell mixing also occurs at the tip of the sprout, leading to tip cell overtaking. This passive cell mixing is in line with the experimental observations of Arima *et al.* [6] and Jakobsson *et al* [5], and arises spontaneously in our models as a side effect of sprouting, without any regulation by Dll4-Notch and VEGF signaling.

We next set out to investigate if Dll4-Notch and VEGF signaling can fine-tune tip cell overtaking in our models when cells have different levels of *Vegfr2* expression. As a first step, we will include Dll4-Notch signaling in our models and study how collective cell behavior during sprouting effects Dll4-Notch patterning (Section 2.4). Subsequently, VEGF signaling is incorporated in the models and simulations will be performed for spheroids that contain a mix of cells with differential levels of *Vegfr2* expression (Section 2.5).

**2.4 Branching, anastomosis and tip cell overtaking affect Dll4-Notch expression**
To study if Dll4-Notch signaling can influence the random tip cell overtaking that we observed in our models, we incorporated a model of the Dll4-Notch signaling network into each of the endothelial cells into both the contact-inhibition and cell elongation models. In this section, we examined how patterning of Dll4 (determining the tip cell phenotype) changes during sprouting, more specifically during branching, anastomosis and tip cell overtaking. To focus on the effect that the local sprout morphology might have on Dll4 patterning, in the simulations presented in this section tip and stalk cells have the same cell behavior, independent of Dll4-Notch activity. In the next section, we will consider differential behavior between tip and stalk cells.

The Dll4-Notch model was based on an ordinary-differential equation (ODE) model proposed by Sprinzak *et al*. [14]. Endothelial cells present Notch receptors and Dll4 ligands on their membranes [2-4]. Upon cell-cell contact, Dll4 ligands activate Notch receptors of neighboring cells through trans-signaling. This activation results in cleavage of Notch and the release of its intracellular domain (NICD). NICD subsequently inhibits the production of Dll4. Notch receptors and Dll4 ligands can also interact and deactivate one another on the same cell, a mechanism that is known as cis-inhibition [14]. To model Dll4-Notch signaling in each cell, each endothelial cell in the model has its own set of ODEs describing the concentration of Dll4, Notch and NICD. To make the level of trans-signaling dependent of the amount of cell-cell contact, the fraction of Dll4 and Notch that a cell presents to an adjacent cell is proportional to the fraction of the cell's membrane that is in contact with it. Cells are assumed to switch between the tip and stalk phenotype when passing a NICD activity threshold: if the NICD level is below the threshold, cells differentiate into tip cells, otherwise they differentiate into stalk cells. The NICD threshold is unknown experimentally; we therefore estimated it such that a salt-and-pepper pattern of alternating tip and stalk cells was formed in agreement with experimental observations [25, 5]. For details on the implementation of tip cell selection, see Section 5.4.

Figures 4A and 4B show that, in agreement with experiments [25, 5], in our models Dll4-Notch signaling generates a checkerboard-like patterning of Dll4. In Figure 4, cells are colored according to a color map, with red representing high levels of Dll4 (tip cells) and blue low levels (stalk cells and extracellular matrix). Also in line with experimental observations [25, 5], cells at the tip position frequently show high concentrations of Dll4. This phenomenon is due to the tip cells' low levels of cell-cell contact with adjacent cells, resulting in a low stimulation of their Notch receptors and, consequently Dll4 production is not inhibited.

Figures 4C-K visualize Dll4-patterning during branching, anastomosis and tip cell overtaking in a simulation of the contact inhibition model, and similar patterns can be seen for the cell elongation model in Figure S3. During branching, new buds are formed and develop over time into growing sprouts, and the leading cell acquires the tip cell phenotype (Figures 4C-E). Figures 4F-H show anastomosis of two sprouts that are led by tip cells. Once the two sprouts meet, they fuse and the two tip cells compete for survival of their tip cell phenotype. Tip cell overtaking is visualized in Figures 4I-K, in which the cell annotated with a star overtakes the cell annotated with a square and subsequently acquires the tip cell phenotype

itself. In summary, branching, anastomosis, and tip cell overtaking induce switching of tip and stalk fates in our models, depending on the relative position, shape and cell-cell contact of the cells in the sprouts.

## 2.5 Effect of VEGF and Dll4-Notch on tip cell overtaking

Our modeling results suggest that tip cell overtaking can occur spontaneously and in absence of Dll4-Notch and VEGF signaling. We next asked how, in our models, Dll4-Notch and VEGF signaling could regulate tip cell overtaking. Jakobsson *et al.* [5] showed in a mosaic sprouting assay using mouse embryonic stem cells that VEGF sensitive cells (wild type, WT) have a higher probability to occupy the tip position than relatively insensitive cells (*Vegfr2* haploid cells, *Vegfr2$^{+/-}$*). After ten days of sprouting, the WT cells occupied 87%, 60% and 40% of the sprout tips when mixed in a 1:1, 1:4 and 1:9 ratio of WT:*Vegfr2$^{+/-}$* cells, respectively. Which mechanisms underlie the increased probability of VEGF sensitive cells to occupy the tip position? We asked whether regulation of cell behavior by VEGF-Dll4-Notch signaling can make VEGF sensitive cells move to the leading position of the sprout.

To address this question, we included a simple model of VEGF signaling into our models: VEGFR2 activation up-regulates Dll4 production, and NICD down-regulates VEGFR2 production [8, 9, 7] (see Section 5.5). *Vegfr2* haploids have half of the VEGFR2 production capacity and therefore have a lower VEGFR2 activity than WT cells. In the *in vitro* experiments of Arima *et al.* [6] and Jakobsson *et al.* [5], VEGF was added uniformly to the growth medium. In our simulation we therefore assumed a uniform field of external VEGF. For simplicity, we will assume in this section that the secreted chemical in the model does not interfere with the external VEGF concentration; i.e. the attractive force is mediated by another chemoattractant (e.g., CXCL12 [26]), by another VEGF-isoform, or even by another means than by chemotaxis (e.g., mechanotaxis [23]).

Tip and stalk cells differ in their behavior, regardless of their genotype. For example, tip cells are more motile than stalk cells and have more VEGF-A-sensitive filopodia, whereas stalk cells proliferate in response to VEGF-A [3]. Tip and stalk cells differentially express genes involved in cell signaling, cell motility and proliferation [27]. We therefore asked which set of differential tip and stalk cell behaviors could cause WT cells to occupy the tip position more often than *Vegfr2* haploids. We first tested if a reduced cell adhesion capacity of tip cells compared to stalk cells can cause VEGF sensitive cells to become sprout leaders, as VEGFR2 activity can cause endocytosis of VE-cadherins and thereby reduce the cell adhesion capacity [28]. To reduce cell adhesion of tip cells in our models, we set the adhesion parameters (J) as follows (with higher values of J giving lower adhesion): $J_{stalk,stalk}=0.2$, $J_{tip,tip}=0.8$, $J_{stalk,tip}=0.8$, $J_{ECM,stalk}=1$, $J_{tip,ECM}=1$. In the contact inhibition model, 93% of the sprout tips in thirty independent simulations were occupied by WT cells for a WT:*Vegfr2$^{+/-}$* ratio of 1:1, 49% for a ratio of 1:4 and 27% for a ratio of 1:9 (Table 1). The results of the 1:1 ratio match the experimental results by Jakobsson *et al.* [5]. WT cells that are located near a sprout tip prefer to become the sprout leader, as the leader cell position has relatively few cell-cell contacts (Figure 5A). The percentages for the lower ratios differ more from the experimental results, because the probability that a WT cell is located near the sprout tip is lower when there are less WT cells in the mix. In this case, WT tip cells manage to go to the outer surface of the sprout, but do not always reach the sprout tip position. In the cell elongation model, the number of WT cells at the sprout tip positions was not significantly different from the number of WT cells at the sprout tips in case of random cell mixing (Table 1). In the cell elongation model, sprout tips often have multiple elongated cells next to each other and a large part of the membrane of the leader cell is in contact with neighboring cells (Figure 5B). The leader cell has much more cell-cell contacts than cells at the sides of the sprout, making it unfavorable for WT tip cells with reduced cell-cell adhesion strengths to become the leader cell in such multi-cellular sprout tips.

Next, we asked if WT cells would more frequently occupy the tip position if the chemoattractant sensitivity differs between tip and stalk cells. Palm *et al.* [29] showed that reduced sensitivity to the chemoattractant increased the potential of a cell to reach the tip position in the contact inhibition model. To further test this hypothesis in our system, we made tip cells less sensitive to the chemoattractant than stalk cells ($\lambda_c$=5 for tip cells and $\lambda_c$=10 for stalk cells), whereas the adhesion energies of tip and stalk cells were set to the same value ($J_{stalk,stalk}$=$J_{tip,tip}$=0.4 $J_{stalk,tip}$=0.4, $J_{stalk,ECM}$=$J_{tip,ECM}$=0.6). Indeed, a reduced sensitivity of tip cells to the chemoattractant compared to stalk cells allowed WT cells to occupy the sprout tip more often than *Vegfr2* haploid cells in the contact inhibition model (ratio WT:*Vegfr2*$^{+/-}$ 1:1 gives a WT tip occupancy of 87%, ratio 1:4 gives 53% and 1:9 gives 34%; Table 1). WT cells are more prone to reach the sprout tip position than *Vegfr2* haploids in the contact inhibition model, because WT cells are less sensitive to the chemoattractant of which the concentration is higher in the sprout center than at the sprout tip as it is secreted by the cells themselves. WT cells do not dominate the tip position in the cell elongation model as strongly as in the contact inhibition model (Table 1). However, the percentage of WT cells at the sprout tips in the cell elongation model is significantly higher than the percentage that would be expected from random cell-mixing. The reduced dominance of WT cells at the sprout tips in the cell elongation model can be explained by the multi-cellular composition of the sprout tips (Figure 5B), as WT cells with a high sensitivity to the chemoattractant are only weakly stimulated to migrate to the tip position in this configuration due to a small difference in concentration of the chemoattractant at the sprout center compared to at the sprout tip.

Thus in our models differential cell behavior of tip and stalk cells can make WT cells occupy the tip position more frequently than *Vegfr2* haploids. In our model, the behavior of *Vegfr2* haploid tip and stalk cells was assumed identical to the behavior of WT tip and stalk cells. What then causes WT tip cells to be overrepresented at the sprout tip relative to *Vegfr2* haploid tip cells? A potential explanation is that WT more easily differentiate to tip cells than *Vegfr2* haploid, due to the higher levels of VEGFR2 and Dll4 in WT cells [5]. To test this possibility, we quantified the number of WT cells and *Vegfr2* haploid cells in the entire cell population (not only at sprout tips) that differentiated into tip cells. Indeed, in our models WT cells are more likely to become tip cell than *Vegfr2* haploids when mixed in a 1:1 ratio and in presence of VEGF. At the end of a simulation of the contact inhibition model, 59 percent of all the WT cells in the population had differentiated into tip cells compared to only 20 percent of the *Vegfr2* haploid cells (percentages measured over *n*=30 simulations). In the cell elongation model, 49 percent of all the WT cells differentiated into tip cells compared to 29 percent of all the *Vegfr2* haploid cells. In conclusion, in our model WT cells have a higher probability to differentiate into the tip cell phenotype than *Vegfr2* haploids as a result of the interactions between VEGFR2 signaling and Dll4-Notch signaling. As a consequence, the tip cells that end up at the tip were more likely to derive from WT cells than from *Vegfr2* haploids.

To study if an external *gradient* of VEGF can affect tip cell overtaking differently than a homogeneous VEGF field, we also performed simulations with the contact inhibition model with differential cell-cell adhesion for tip and stalk cells in the presence of an external VEGF gradient. We only let tip cells chemotact towards VEGF to simulate the most extreme advantage for tip cells. The presence of a VEGF gradient rather than a uniform VEGF field did not significantly change the mean tip cell overtake frequency in sprouts (Figure S4), the sprout tip occupancy by WT versus *Vegfr2*$^{+/-}$ cells (Table S1) or the cell trajectory analysis results (Table S2). Once VEGFR2 is stimulated by VEGF, lateral inhibition by Dll4-Notch signaling quickly generates a comparable alternating tip-stalk pattern as in the presence of a uniform VEGF field.

In conclusion, simulation results of the contact inhibition model suggest that VEGF-Dll4-Notch signaling might tune which cells ends up at the sprout tip position when cells have different levels of *Vegfr2* expression. To make this possible tip and stalk cells must behave differently, such as differential cell-cell adhesion or differential sensitivity to an attractant. Interestingly, in the cell elongation model *Vegfr2* expression did not significantly affect the ability of cells to reach the tip cells position. The multi-cellular sprout tip environment is unfavorable for cells with such cell behaviors, suggesting that sprout morphology can affect the regulation by VEGF and Dll4-Notch signaling in tip cell overtaking.

## 3. Discussion

Our simulation results show that the collective cell behavior responsible for *in silico* angiogenesis-like sprouting produces cell mixing and tip cell overtaking dynamics in accordance with published measurements [6]. The contact inhibition model as well as the cell elongation model reproduced the experimental results of Arima *et al.* [6], who found that tip and stalk cells mix at sprout tips. Our modeling results thus show that tip cell overtaking can occur *as a side effect of sprouting* and might not be necessarily functional.

In disagreement with this conclusion but in agreement with Jakobsson *et al.* [5], in the contact inhibition model the activity of the VEGF-Dll4-Notch signaling network affected the competitiveness of cells for the tip cell position. A possible interpretation is that tip cell overtaking is genetically regulated, implying that tip cell overtaking must be functional. Jakobsson *et al.* [5] proposed that tip cell overtaking allowed for the most VEGF sensitive cell to become the leader cell at all times to optimally respond to VEGF in the environment. Alternatively, based on our modeling results that suggest that tip cell overtaking occurs as a side effect of sprouting, we propose that the VEGF-Dll4-Notch signaling network makes the cell in the tip position cross-differentiate into a tip cell. Here the VEGF-Dll4-Notch signaling network would act to *protect* the growing sprout against the loss of a tip cell at the sprout front due to random cell mixing. In this interpretation tip cell overtaking would be a purely random side effect of sprouting and be non-functional in itself.

Our simulations also suggest that the morphology of the sprout tip might be important to tip cell overtaking. The sprout tip position was less favorable for tip cells with reduced cell-cell adhesion or reduced sensitivity to the chemoattractant in the cell elongation model: sprouts in the cell elongation model consist of multiple cells parallel to one another, whereas in the contact inhibition model (and in many actual angiogenic sprouts) only one cell leads the sprouts.

Bentley *et al.* [9] assume in their model that long-range cell movements during cell mixing are driven by Notch/VEGFR-regulated differential dynamics of VE-cadherin junctions. Their simulations suggest that the observations by Jakobsson *et al.* [5] are best reproduced when tip cells have a reduced cell-cell adhesion compared to stalk cells, and are more polarized than stalk cells, preferentially extending protrusions towards the sprout tip. In contrast to the results by Bentley et al., in our simulations, cell mixing occurs spontaneously without any assumptions on differential adhesion or polarization. This discrepancy could be caused by a difference in the models. Whereas in the model of Bentley *et al.* [9] cells can only migrate relative to a static sprout, in our models sprout formation emerges from the assumptions on cell behavior. In simulations with the contact inhibition model, differential cell-cell adhesion between tip and stalk suffices to reproduce the results by Jakobsson *et al.* [5]. Because sprout extension biases cell movement towards the tip, we do not require explicit tip-directed cell polarization.

Although the contact inhibition variant of our model best reproduced the experimental observations on tip cell overtaking, our previous motivation for assuming contact inhibition of chemotaxis is inconsistent with the present model. We previously assumed that contact-dependent phosphorylation of VEGFR2 by VE-cadherin mediates contact-inhibition of chemotaxis [16, 13]. Recent work showed that VEGFR2 activity internalizes VE-cadherins [28]. If this mechanism were implemented in our model, high VEGFR2 activity in the tip cells would internalize VE-cadherins and reduce the strength of VE-cadherin-mediated contact inhibition. The chemotactic sensitivity to VEGF in these tip cells would thus increase and tip cells would move towards the center of the sprouts, inconsistent with biological observations. Potential fixes for this experimental discrepancy include (a) the possibility that cells do not aggregate via VEGF, but via another chemoattractant or attractive forces, or (b) to consider matrix-bound VEGF [20, 21] in our models, which would only be available at the periphery of the spheroids.

By what mechanisms are cells driven forwards and backwards along sprouts? Apart from the random cell motility the Cellular Potts model describes, the chemoattractant gradients seem to play a key role in our model. The models predict that the concentration of chemoattractant will be higher in the center of the sprout than at the flanks, and higher in concave regions of the sprout surface than at convex regions. Cells in the center of the sprout are, therefore, constrained by the gradient, whereas a compressive force towards the center of the sprout pushes the cells forwards. Cells on the flank of the sprout sense a shallower gradient and are therefore more motile, allowing them to walk backwards along the sprout towards the high concentration of the chemoattractant at concave branch points. Experimentally, it will be interesting to validate this hypothesis by comparing the relative position of cells in the sprout to the migration direction within the sprout. Besides by a chemoattractant, the attractive force could be caused by other biological mechanisms, such as mechanical strains in the extracellular matrix [23] or signaling through long filopodia [30]. In our ongoing research we are investigating whether forward and backward motion indeed requires a chemotactic gradient or if it can also be driven by other mechanisms such as cell-cell adhesion [17] or mechanotransduction via the ECM [23].

# 4. Conclusions

Tip cell overtaking has been studied in different experimental setups [6, 5], but the biological function is still unknown. We asked whether tip cell overtaking is merely a side effect of sprouting or whether it is regulated through a VEGF-Dll4-Notch signaling network, and thus might be functional. For this purpose, we studied two existing computational models of angiogenic sprouting, allowing us to study the effect of sprouting dynamics on tip cell overtaking. In our models, cells spontaneously move back and forth along the sprout as a side effect of the sprouting mechanisms, as was seen in experiments of Arima *et al.* [6]. This suggests that tip cell overtaking and sprouting dynamics may be interdependent and, therefore, should be studied and interpreted in combination. In experiments with mosaic endothelial spheroids [5], it was found that wild type cells have a competitive advantage over *Vegfr2* haploid cells for the tip cell position, suggesting that VEGF-Dll4-Notch signaling regulates tip cell overtaking. In agreement with these experiments, in one of our models the wild type cells also end up at the tip position more frequently than *Vegfr2* haploids, simply because the wild type cells more often differentiate into tip cells. This would suggest that VEGF-Dll4-Notch signaling can regulate tip cell overtaking. Based on the model results that tip cell overtaking is a non-functional *side effect* of sprouting, we suggest an alternative function for VEGF-Dll4-Notch signaling: Rather than regulating which cell ends up at the tip, it might assure that the cell that randomly ends up at the tip position acquires the tip cell phenotype.

# 5. Methods
## 5.1 Angiogenesis models
To model angiogenic sprouting [12, 13], we made use of a modified Cellular Potts, a widely used, cell-based simulation technique. Although other modeling techniques have been used to model angiogenesis, including continuum approaches [31, 32, 24] and single-particle cell-based techniques based on Lagrangian dynamics [33, 18, 34], in this study it was crucial to follow the trajectories of individual cells and to allow cells to assume flexible cell shapes. We therefore made use of a multi-particle, cell-based model, a class of cell-based simulation techniques in which one cell is represented by a collection of lattice sites [35]. Among this class of models, the Cellular Potts model [36, 37] is a widely used and computationally efficient technique, which has been used to study *de novo* angiogenic sprouting sprouting [12, 13, 38-40].

**Cellular Potts Model**

In the CPM, cells are projected on a regular square lattice $\Lambda \subset \mathbb{Z}^2$. The cells are represented as patches of connected lattice sites $\vec{x}$, with each site of a cell having the same cell identifier, $\sigma(\vec{x}) \in \mathbb{N}$. Lattice sites not occupied by cells belong to extracellular matrix (ECM) with $\sigma = 0$. A further identifier, $\tau(\sigma) \in \{\text{tip}, \text{stalk}\}$, differentiates the tip and stalk cells. A Hamiltonian energy ($H$) gives the force balance following from the properties and behaviors of the cells,

$$H = \sum_{(\vec{x},\vec{x}')} J(\tau(\sigma(\vec{x})), \tau(\sigma(\vec{x}')))(1 - \delta(\sigma(\vec{x}), \sigma(\vec{x}'))) + \lambda_{size} \sum_\sigma ((A(\sigma) - a(\sigma))^2 + H'. \quad (0.0)$$

Here $J$ represents the interfacial energies between the cells, due to cell-cell adhesion and cortical tensions [41]; the Kronecker-delta construction ($\delta(x, y) = \{1, x = y; 0, x \neq y\}$) selects the cell-cell interfaces. The second term constrains the volumes of the cells (or areas in this two-dimensional model), with $A(\sigma)$, the resting area and $a(\sigma)$ the actual area of the cell. Further constraints, used to represent additional cell behaviors, are including in the third term, $H'$. These are defined in the next sections.

The cells move by attempting to extend or retract pseudopods, which are mimicked by copying the state ($\sigma(\vec{x})$) of a randomly selected lattice site into a randomly selected adjacent lattice site $\vec{x}'$. A copy that reduces the Hamiltonian represents a move along a force and is always accepted. To represent active surface fluctuations (generated by actin dynamics) a copy that increases the Hamiltonian is accepted according the Boltzmann probability function: $P_{Boltzmann}(H) = e^{\frac{-\Delta H}{\mu}}$, with $\mu$, the random, active cell parameter; throughout this paper, we set $\mu = 1$. Time is measured in *Monte Carlo Steps* (MCS), where one MCS represents as many copy attempts are performed as there are sites in the lattice. One MCS corresponds to thirty seconds.

**Cell elongation**

To constrain the cell length ($l$) in the cell elongation model, an additional constraint is used as previously described [12]. Briefly, $H_{length} = \lambda_{length}(\sigma) \sum_\sigma ((L(\sigma) - l(\sigma))^2$, with $\lambda_{length}(0) = 0$ and $\lambda_{length}(\sigma) > 0$ for all $\sigma > 0$, i.e., the length constraint holds for the cells only. $L(\sigma)$ and $l(\sigma)$ are the target cell length and current cell length. The current cell length can be efficiently estimated from the cell's inertia tensor, as described previously. To prevent cells from splitting up in an attempt to optimize the moments of inertia, a large penalty ($H_{\text{connectivity}}$) is added to the Hamiltonian in case a copy would split up a cell locally.

**Chemoattractant secretion**
We assume that the endothelial cells secrete a chemical signal, $c(\vec{x})$, which diffuses and degrades according to a partial-differential equation (PDE) coupled to the CPM,

$$\frac{\partial c}{\partial t} = \alpha(1-\delta(\sigma(\vec{x}),0)) - \varepsilon\delta(\sigma(\vec{x}),0)c + D\vec{\nabla}^2 c.  \qquad (0.0)$$

The cells secrete the signal at rate $\alpha$ per second, it is degraded at a rate $\varepsilon$ per second, and it diffuses in the ECM at rate $D$ m$^2$/s. The Kronecker-delta constructions indicate that the cells secrete the chemoattractant, which is degraded in the ECM ($\delta(\sigma(\vec{x}),0)=0$ is inside cells and $\delta(\sigma(\vec{x}),0)=1$ in the ECM). After each MCS, this partial differential equation is solved numerically using a finite-difference scheme on a lattice that matches the CPM lattice, using 15 diffusion steps per MCS with Δt=2 s and Δx=2μm.

**Chemotaxis**
To model chemotaxis, we bias the update probabilities such that membrane fluctuations up gradients of the chemoattractant are [42] favored. To this end, we modify the Hamiltonian during each copy attempt, $\Delta H_{\text{chemotaxis}} = \Delta H + \lambda_c(c(\vec{x}) - c(\vec{x}'))$, with $\lambda_c$ a parameter giving the sensitivity to the chemoattractant. The contact inhibition model assumes that cell-cell contact inhibits chemotaxis: i.e., $\lambda_c$ becomes zero for copies at cell-ECM interfaces.

**Model set up**
The contact inhibition model [13] and the elongation model [12] make use of the standard Cellular Potts model, and the chemoattractant diffusion and chemotaxis models, where the contact inhibition model restricts chemotaxis to cell-matrix interfaces as described above. The cell elongation model additionally includes a cell length constraint. The simulations are initialized with a spheroid of cells, of radius of 45 lattice sites containing square cells of 7 lattice sites wide, surrounded by extracellular matrix. The simulations are initiated with cell spheroids. In these models, sprout form after 30000 MCS, corresponding to approximately ten days of sprouting. At 10000 MCS we start to monitor tip cell overtakes and cell mixing in the models. The parameter values for both models, obtained from [12, 13], are listed in Table S3. The models were implemented with the modeling environment CompuCell3D, scripts are available on request.

**5.2 Leader cell identification**
To identify leader cells in a network of endothelial cells, sprouts are detected by converting the network of cells into a graph of edges, branch nodes and end nodes as in [12]. To this end, the irregularities of the network are closed with a morphological closing operation using a disk of radius (*r*), the network is thinned by a radius (*t*) and subsequently the branches are pruned with a distance (*p*) [43]. Nodes within a range of *m* lattice sites are merged. The settings to create graphs from simulated networks in the contact inhibition model are *r*=4, *t*=4, *p*=10, and *m*=10, and for the cell elongation model *r*=2, *t*=5, *p*=25, and *m*=15. A sprout is defined as a connection between a branch point *B* and an endnote *E*.

     The leader cell of a sprout is found in a few steps. The first guess (*G*) for the leader cell is the cell in which the endnote *E* is located. If *E* happens to be located in the ECM, the

cell belonging to the most frequently occurring cell identifier in the set of neighboring lattice sites of *E* is selected as *G*. Next, a straight line (*e*) is drawn from *B* through *E* in the direction of the sprout tip. The furthest cell lattice site on this line in the sprout, after which at least five consecutive ECM lattice sites follow, is identified as *T*. Subsequently, a line (*a*) perpendicular to the line *e* and through *T*, is constructed (Figure 6). All cells on line *a* that are neighbors of cell *G* become additional candidates for leader cell. Each of these cells that are connected to node *B* through at least an equal amount of cells as *G* is, taking the shortest path according the Dijkstra algorithm through a graph in which each cell is a node and shares an edge with the node belonging to a neighboring cell, remain candidate together with cell *G*. The cell that has the lattice site with the largest distance to *B* (indicated with a star in Figure 6) becomes the leader cell of the sprout.

## 5.3 Cell trajectory analysis

Cells are tracked during a simulation by storing the position of their center of mass every 20 MCSs. This cell trajectory data is used to calculate cell coordination and directional motility by the methods described by Arima *et al.* in [6]. Two adaptations have been made compared to the methods used by Arima *et al.* [6] to automate the analysis: defining a sprout and defining the elongation axis of a sprout. We define a sprout as the leading cell (see Methods Section 5.2) together with its ten nearest neighbors in the same sprout. The ten nearest neighbors are found by listing the cells that contact the leader cell and subsequently listing the cells they contact that are not listed yet and so on, until ten cells are listed. We defined the elongation axis as the edge between the start and end position of a sprout. The start position is the average of the position of the branch node at the first and last time frame of the existence of a sprout. The end position is the average of the tip position for these two time frames. This was required since sprouts often shift and curve. Cell coordination and directional motility are calculated according to the methods in Arima *et al.* [6]. We have averaged the results over the sprouts (or the cells in the sprouts) formed during 15 simulations with different random seeds. In the calculation for the directional motility, cells that traveled a smaller distance than 0.5 lattice sites [6] are considered to be stopped. The dispersion coefficient of cells during sprouting can be derived from the mean square displacement ( $\text{MSD} = <(\vec{x}(0) - \vec{x}(t))^2>$ ) of the centers of mass of all cells within sprouts measured each 20 MCS during sprouting time, with the data of all 15 simulations grouped. For this purpose, we measured the one-dimensional displacement of the projection of the centers of mass of cells on the sprouting elongation axis. The dispersion coefficient ( $D$ ) and the sprout elongation velocity ( $v$ ) are derived by fitting the MSD curve with $<(\vec{x}(0) - \vec{x}(t))^2> = 2Dt + (vt)^2$.

## 5.4 Dll4-Notch signaling model

A model of lateral inhibition by Dll4-Notch signaling is included in each cell of the CPM. The model is based on an ordinary-differential equation (ODE) model previously proposed by Sprinzak *et al.* [14]. In this model, Notch binds Dll4 ligands in adjacent cells (*trans*-interaction) leading to the production of NICD; Notch and Dll4 also bind intracellularly leading to inhibition of NICD production. Such *cis*-inhibition makes the Dll4 and Notch lateral inhibition mechanism more robust to noise [14] and has been observed, e.g., in the Drosophila wing [44] and eye [45]. *Cis*-inhibition of Dll4 and Notch remains to be confirmed in endothelial cells; recent modeling work [46] suggests, however, that it has little effect on the robustness of tip cells.

The model is described by the following set of ODEs:

$$\frac{dS_i}{dt} = \alpha_S \frac{\left(\frac{1}{d^2}\sum_{j\in NB(\sigma)} N_i D_j \frac{|P_{i,j}|^2}{|P_i||P_j|}\right)^{n_S}}{k_S + \left(\frac{1}{d^2}\sum_{j\in NB(\sigma)} N_i D_j \frac{|P_{i,j}|^2}{|P_i||P_j|}\right)^{n_S}} - \gamma_S S_i \tag{0.0}$$

$$\frac{dD_i}{dt} = \beta_{Dc} + \frac{\beta_D}{1+S_i^{m_D}} - \gamma_D D_i - \frac{D_i N_i}{k_c} - \frac{1}{k_t d^2}\sum_{j\in NB(\sigma)} D_i N_j \frac{|P_{i,j}|^2}{|P_i||P_j|} \tag{0.0}$$

$$\frac{dN_i}{dt} = \beta_N - \gamma_N N_i - \frac{N_i D_i}{k_c} - \frac{1}{k_t d^2}\sum_{j\in NB(\sigma)} N_i D_j \frac{|P_{i,j}|^2}{|P_i||P_j|} \tag{0.0}$$

Each cell $i$ has an individual concentration of Dll4 ($D_i$), Notch ($N_i$) and activated Notch signal ($S_i$) representing NICD. The ODE model contains constants for constitutive production of Notch and Dll4 ($\beta_N$ and $\beta_{Dc}$), decay constants for Notch ($\gamma_N$), Dll4 ($\gamma_D$) and NICD ($\gamma_S$), a cis-interaction coefficient ($k_c$), a trans-signaling coefficient ($k_t$) and a scaling factor ($d$). Trans-signaling results in NICD production following a Hill equation ($n_S, k_S$), with a production rate ($\alpha_S$). The variable Dll4 production ($\beta_D$) is inhibited by NICD using a repressive Hill function ($m_D$). In contrast to the Sprinzak model, our model considers the size of cell-cell contacts for trans-signaling. Dll4 and Notch are assumed to be spread homogeneously over all lattice sites in the membrane of the cell ($P_i$). Cell $i$ and neighboring cell $j$ contact each other at region $P_{i,j}$ of the cell membrane. Cell $i$ will present a fraction of its Dll4 receptors to its neighbor, proportional to the length of the contacting cell membrane region ($|P_{i,j}|$) divided by the total length of the membrane ($|P_i|$). This results in contact-surface dependent trans-signaling obeying: $D_i(|P_{i,j}|/|P_i|) * N_j(|P_{j,i}|/|P_j|)$. The collection of cells that are in contact with cell $i$ are represented by the set $NB(\sigma)$. We solve these equations ten times per MCS with Δt=3 s. The reference parameter values of the model by Sprinzak *et al.* [14] were rescaled after the extension of the contact-surface dependent trans-signaling to obtain the experimentally observed tip and stalk patterns as discussed in Section 2.4. The parameter values of the Dll4-Notch signaling network are listed in Table S4.

### 5.5 Modeling of Dll4-Notch signaling in presence of VEGF

VEGF signaling was added to the tip cell selection model described in Section 5.4. A non-diffusive, constant, homogeneous, external VEGF ($V$) field with a value of one was added to the model. The equations that are altered or added due to the presence of VEGF relative to the Dll4-Notch signaling equations (Section 5.4) are:

$$\frac{dR_i}{dt} = \beta_{Rc} + \frac{\beta_R}{1+S_i^{m_R}} - \gamma_R R_i \tag{0.0}$$

$$\frac{dA_i}{dt} = \alpha_A \frac{\left(\sum_{j \in P_i} \frac{R_i V_j}{|P_i|}\right)^{n_A}}{k_A + \left(\sum_{j \in P_i} \frac{R_i V_j}{|P_i|}\right)^{n_A}} - \gamma_A A_i \tag{0.0}$$

$$\frac{dD_i}{dt} = \beta_{Dc} + \frac{\beta_D}{1+S_i^{m_D}} - \gamma_D D_i - \frac{D_i N_i}{k_c} - \frac{1}{k_t d^2} \sum_{j \in NB(\sigma)} D_i N_j \frac{|P_{i,j}|^2}{|P_i||P_j|} + \alpha_D \frac{A_i^{n_D}}{k_D + A_i^{n_D}}. \tag{0.0}$$

The equations for solving $N_i$ and $S_i$ remain the same, and two equations are added that describe the VEGFR2 concentration ($R_i$) and the VEGF signaling activity ($A_i$) of cell i. The total VEGF concentration a cell perceives at its membrane lattice sites ($\sum_{j \in P_i} \frac{R_i V_j}{|P_i|}$) upregulates its VEGF signaling activity with production rate $\alpha_A$, following a Hill equation ($n_A, k_A$). VEGF signaling activity has a decay constant ($\gamma_A$) and VEGFR2 has a decay constant ($\gamma_R$). An additional term is present for Dll4 that expresses the positive feedback of VEGF activity on the Dll4 production, modeled with a Hill equation ($n_D, k_D$) and a production rate ($\alpha_D$). *Vegfr2*[+/-] cells are modeled by multiplying the constant production of VEGFR2 ($\beta_{Rc}$) and the variable production ($\beta_R$), which is inhibited by NICD ($S_i$) using a repressive Hill equation ($m_R$), by a half. The parameter values of the VEGF-Dll4-Notch signaling network are listed in Table S4. We manually fitted the parameters for VEGF-Dll4-Notch signaling, such that the experimentally observed tip and stalk patterns (as discussed in Section 2.4) are maintained, and in addition, that Dll4 and VEGFR2 levels are correlated with one another as shown by Jakobsson *et al.* [5].

## Competing interests
The authors declare that they have no competing interests.

## Authors' contributions
SB and RM designed the study and drafted the manuscript. SB performed the simulations and analyzed the data. All authors have read and approved the final version of the manuscript.

## Acknowledgements
We thank Indiana University and the Biocomplexity Institute for providing the CompuCell3D modeling environment and SURFsara (www.surfsara.nl) for the support in using the Lisa Compute Cluster. The investigations were supported by the Division for Earth and Life Sciences (ALW) with financial aid from the Netherlands Organization for Scientific Research (NWO).

# Figures

**Figure 1: Overview of the workflow.**
We studied the biological relevance and the driving mechanisms of tip cell overtaking. (A) As a first step, we asked whether tip cell overtaking can be a side effect of sprouting. We studied tip cell overtaking in two computational models of angiogenic sprouting (the contact inhibition model and cell the elongation model), with different sprouting dynamics. We quantified tip cell overtaking and cell kinetics during simulations of these models and compared the results with similar *in vitro* experiments of Arima et al. [6]. (B) As a next step, we asked if tip cell overtaking can be regulated by VEGF-Dll4-Notch signaling. We added a VEGF-Dll4-Notch signaling network to each cell in the two models of angiogenic sprouting. Simulations are initialized with spheroids that contain a mix of wild type (WT) cells and $Vegfr2^{+/-}$ cells. Due to signaling, cells can switch between four phenotypes during sprouting: WT tip cell, WT stalk cell, $Vegfr2^{+/-}$ tip cell, and $Vegfr2^{+/-}$ stalk cell. At the end of the simulations we quantified the percentage of sprout tips that were occupied by WT cells and compared the simulation results to experimental results of Jakobsson *et al.* [5].

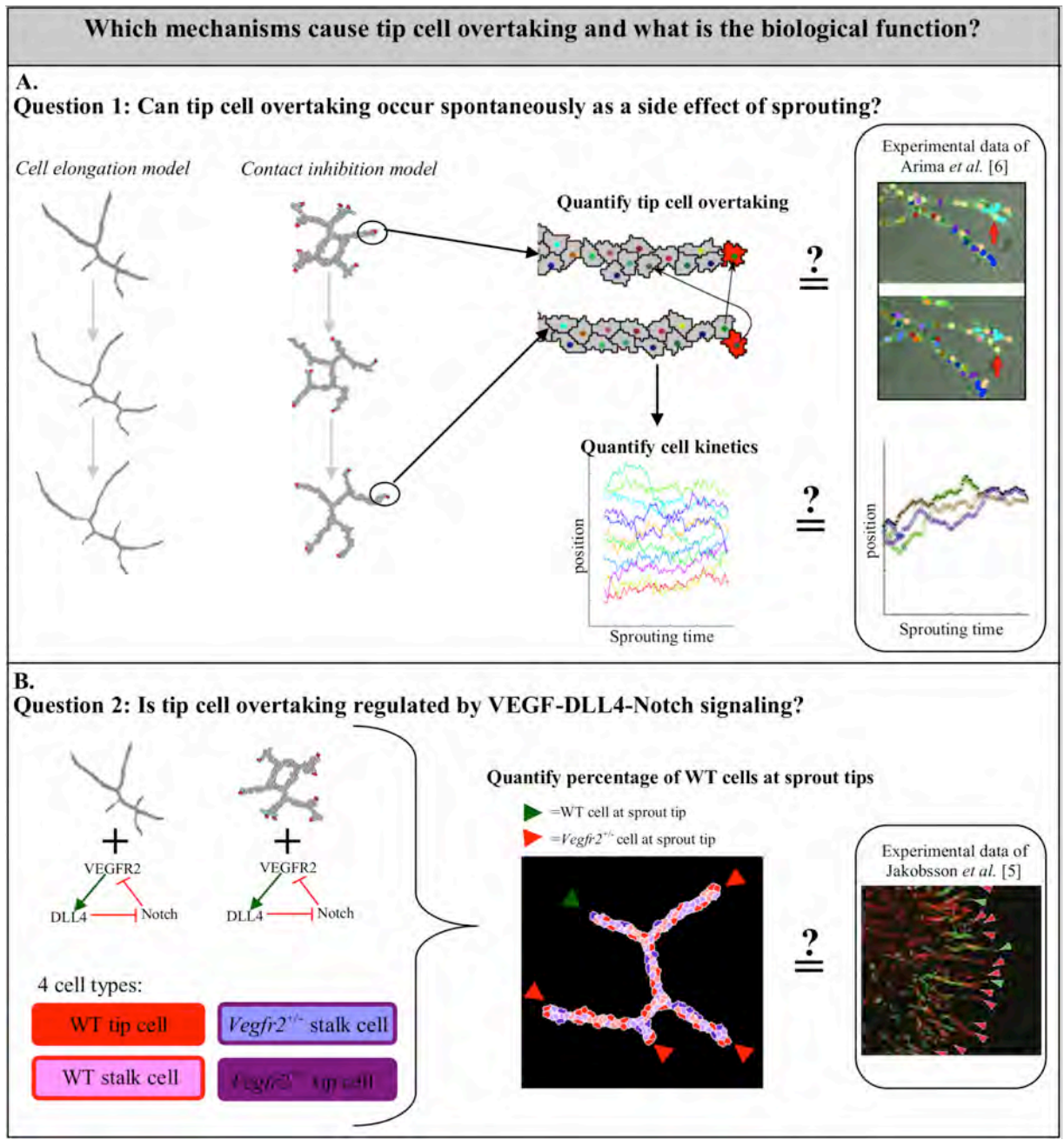

**Figure 2: Leader identification and tip cell overtaking in the contact inhibition and cell elongation model.**
Sprouts formed from a spheroid in 30000 MCS by (A) the contact inhibition model and by (B) the cell elongation model. Red cells at the sprout tips indicate the identified leader cells. Tip cell overtaking occurs in the (C) contact inhibition model as well as in (D) the cell elongation model. Two images of the same sprouts are shown for each model, with the lower sprout being at a later time point than the upper sprout. The center of mass is depicted with a colored dot for each cell and the displacement of the leader cells in time is visualized with the arrows. The mean tip cell overtake rate per sprout, calculated over 15 independent stochastic simulations, is 0.67 (±1.32) overtakes per 20000 MCS for the contact inhibition model and 4.59 (±5.24) overtakes per 20000 MCS for the cell elongation model.

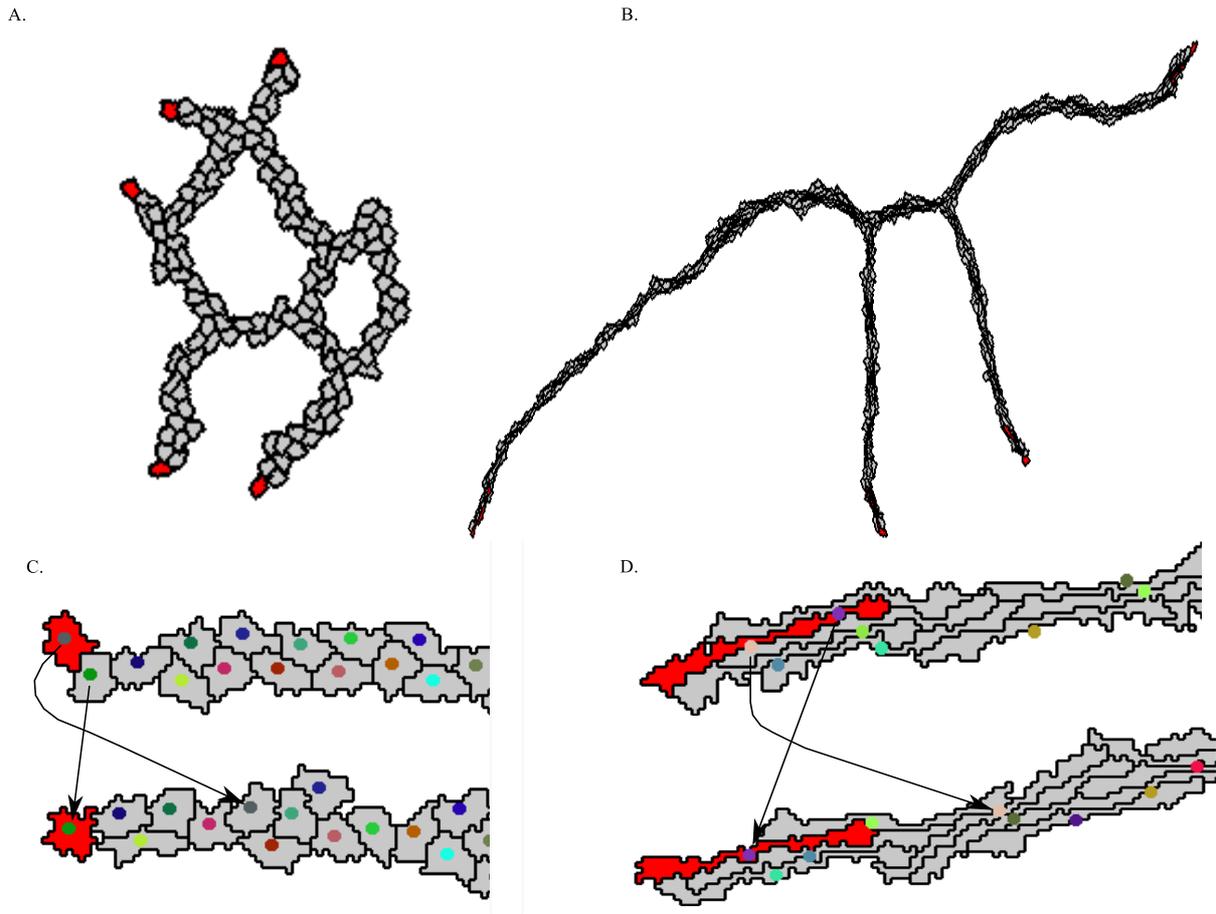

**Figure 3: Analysis of cell migration within sprouts.**
The position of each cell is orthogonally projected onto the sprout elongation axis and plotted against sprouting time in minutes for (A) a sprout in a murine aortic ring assay (Figure A is adapted from [6]), (B) in the contact inhibition model and (C) in the cell elongation model; arrows indicate tip cell overtake events. The standard deviation std(θ/π) is given for (D) anterograde moving cells (θ<π/2) and (E) retrograde moving cells (θ>π/2) for the experimental observations by Arima *et al*. [6] (exp), for the contact inhibition model (contact) and for the cell elongation model (long). (F) Directional motility represents the percentage of cells moving anterograde (blocked pattern), retrograde (diagonal striped pattern) or stopped (horizontally striped pattern). Mean square displacement (MSD) of cells, calculated by the projection of the center of mass on the sprout elongation axis, plotted against sprout time for (G) the contact inhibition model and for (H) the cell elongation model. The fluent blue line represents the fitted curve following: $\text{MSD} = 2Dt + (\text{v}\,t)^2$, with $D$ the dispersion coefficient and $v$ the sprout elongation velocity.

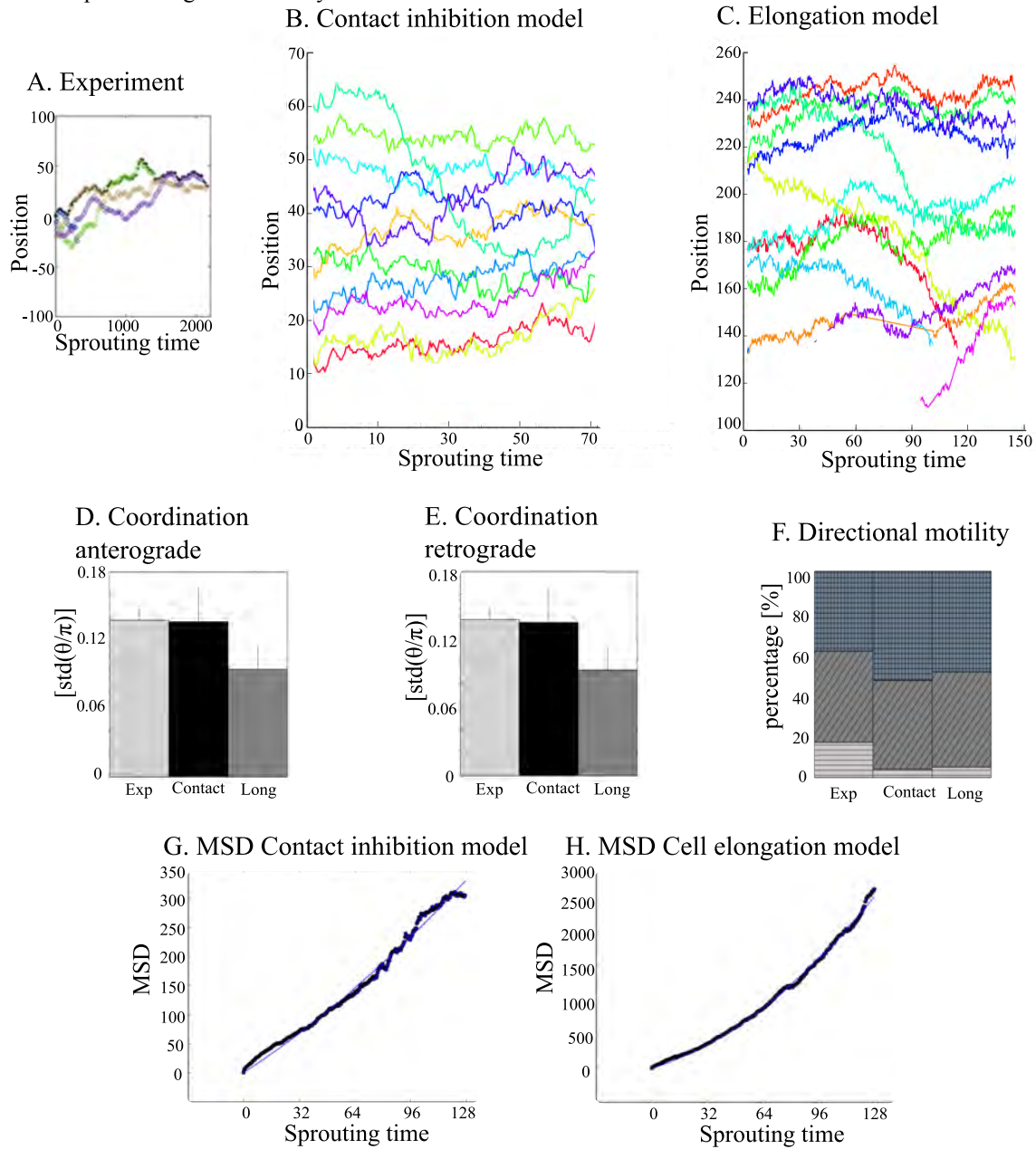

**Figure 4: Dll4 patterning by tip cell selection.**
(A) Checkerboard-like patterning of tip and stalk cells in a simulation of the contact inhibition model. The red color indicates high levels of Dll4 (tip cells) and blue indicates low levels of Dll4. (B) Checkerboard-like patterning of tip and stalk cells in a simulation of the cell elongation model. Figures C-J are images from a simulation of the contact inhibition model. (C-E) Enlarged view of a sprout in which branching occurs over time, at the location of the white circle in panel C. (F-H) Enlarged view of two fusing sprouts (anastomosis) in time, indicated by the white circle in panel F. (I-K) Enlarged view of a sprout in which tip cell overtaking occurs in time at the location of the white circle in panel I. The cell annotated with a square overtakes the tip cell position from the cell annotated with a star.

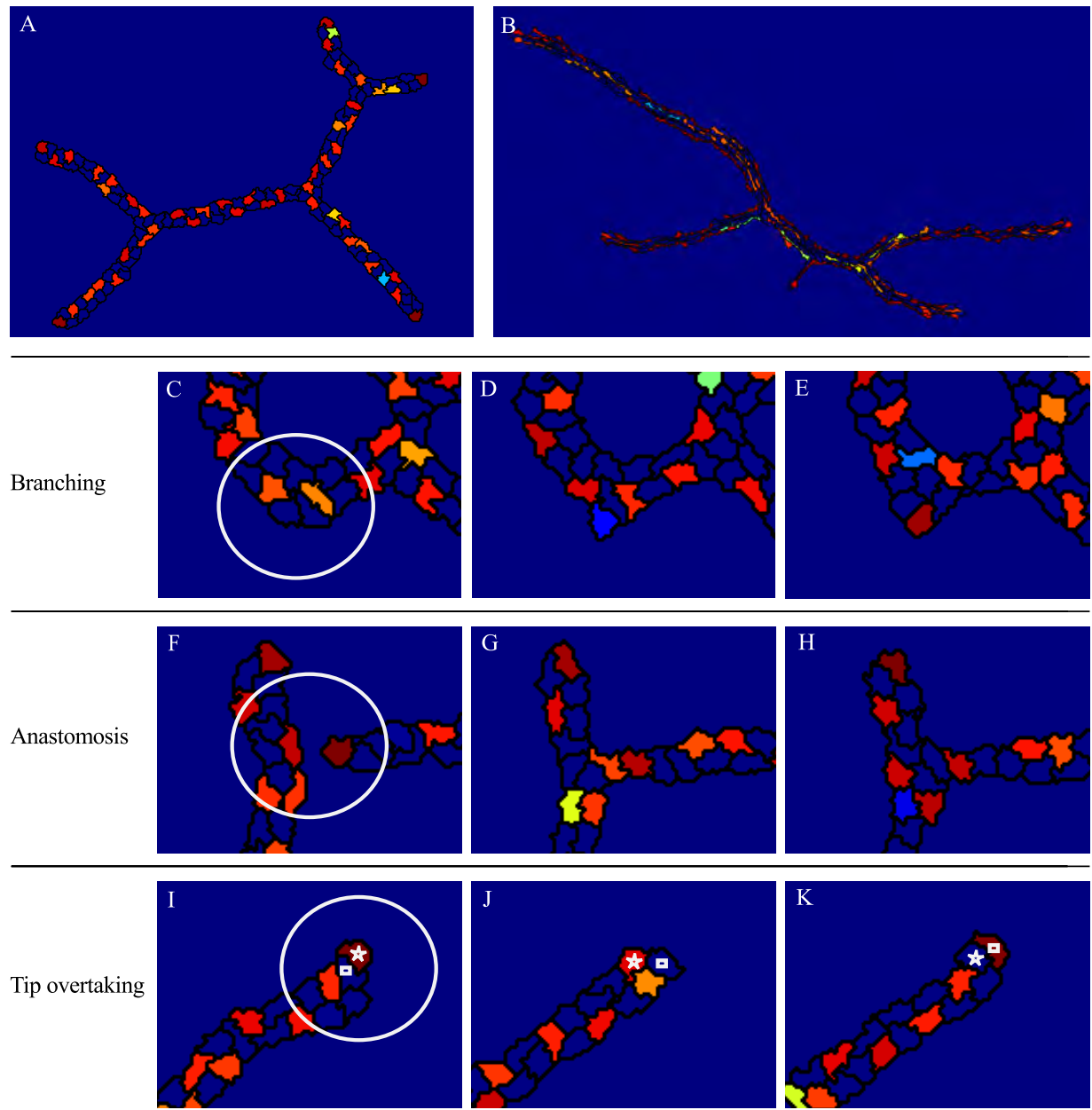

**Figure 5: Relative cell positions at sprout tips.**
Enlarged view of a sprout tip in a simulation of (A) the contact inhibition model and of (B) the cell elongation model. WT tip cells are colored red, *Vegfr2* haploid tip cells dark purple and *Vegfr2* stalk cells light purple. The leader cells of the sprouts are marked with yellow stars. The leader cell is the contact inhibition model has relatively little cell-cell contact compared to other cells in the sprout, while the leader cell in the cell elongation model is in contact with other cells for a large part of its membrane due to the multi-cellular composition of the sprout tip.

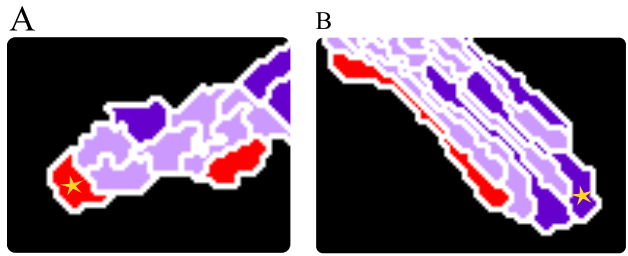

**Figure 6: Leader cell identification.**
Schematic representation of a sprout to illustrate the identification of the leader cell. Line *e* is drawn through nodes *B* and *E* to find *T*, the furthest lattice site in the sprout on line *e*. Line *a* is perpendicular to line *e* and through *T*. The cell in which *E* is located and its neighbors that are on line *a*, are candidates to become the leader cell. The cell with the lattice site farthest from *B* (indicated with a star) and is connected to *B* through at least an equal amount of cells, will become the leader cell (indicated in red).

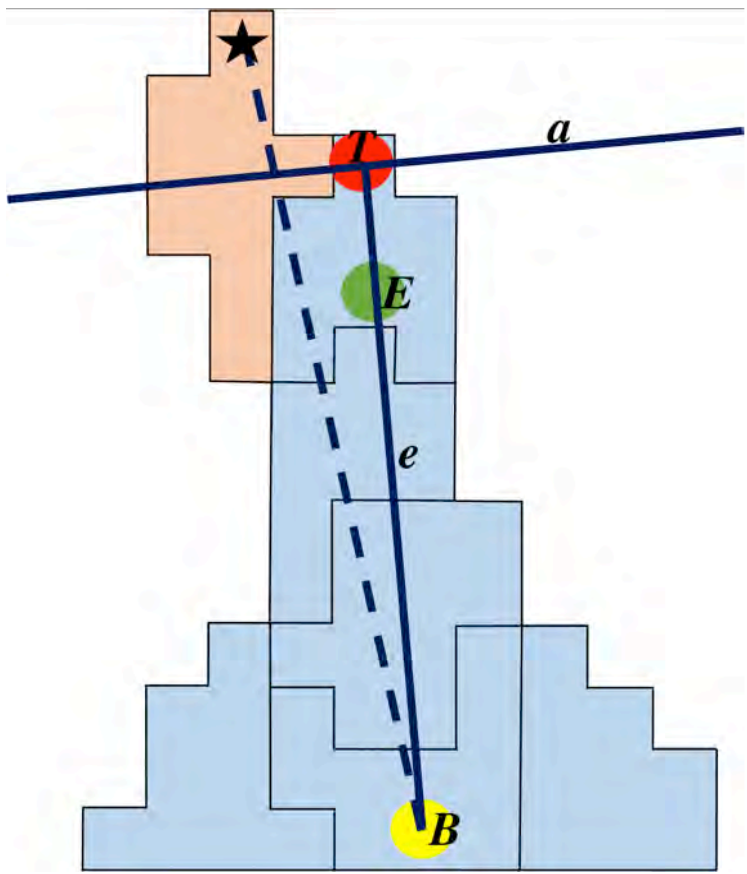

# Tables

**Table 1: Sprout tip occupancy by WT cells.**
Overview of the percentile sprout tip occupancy by WT cells. WT occupancy was quantified for different initial WT:$Vegfr2^{+/-}$ mixing ratios in experiments [5] (Experiment), in the contact inhibition model (Contact) and in the cell elongation model (Long). The WT:$Vegfr2^{+/-}$ mixing ratios were 1:1, 1:4 and 1:9, resulting in a WT percentage of 50, 20 and 10 respectively. Two different mechanisms are tested in the models: differential adhesion between tip and stalk cells and differential sensitivity to an auto-secreted chemoattractant between tip and stalk cells. The p-values represent the probability that the total number of simulated sprouts were occupied by at least the indicated percentage of WT cells when assuming only random motion (calculated with a binomial distribution, with n the number of sprouts, k the number of sprouts occupied by WT cells, and p the mixing ratio).

|               |            | Differential adhesion |                | Differential sensitivity to chemoattractant |                |
| ------------- | ---------- | --------------------- | -------------- | ------------------- | -------------- |
| WT percentage | Experiment | Contact               | Long           | Contact             | Long           |
| 50            | 87         | 93 (p=6.7·$10^{-16}$) | 48 (p=0.73)    | 87 (p<1·$10^{-16}$) | 64 (p=6.2·$10^{-4}$) |
| 20            | 60         | 49 (p=7.7·$10^{-16}$) | 18 (p=0.75)    | 53 (p<1·$10^{-16}$) | 25 (p=3.6·$10^{-2}$) |
| 10            | 40         | 27 (p=6.9·$10^{-9}$)  | 11 (p=0.33)    | 34 (p<1·$10^{-16}$) | 22 (p=7.2·$10^{-8}$) |

# Supplementary files

**Movie S1: Tip cell overtaking in the contact inhibition model.**
Tip cell overtakes are visible during sprouting in a simulation of the contact inhibition model. The center of mass of each cell is depicted with a colored dot to allow tracking of individual cells.

See: https://youtu.be/9cvQ3zA3_b4

**Movie S2: Tip cell overtaking in the cell elongation model.**
Tip cell overtakes are visible during sprouting in a selected sprout in a simulation of the cell elongation model. The center of mass of each cell is depicted with a colored dot to allow tracking of individual cells.

See: https://youtu.be/S8oX5xXgagQ

**Figure S1: Sensitivity of tip cell overtaking in contact inhibition model.**
The mean overtake rate per sprout, based on 15 independent simulations, is plotted against cell-cell adhesion ($J_{cell,cell}$) and cell-ECM adhesion ($J_{cell,ECM}$), sensitivity to the auto-secreted chemoattractant ($\lambda_c$), the cellular temperature ($\mu$), the diffusion constant of the chemoattractant (D), the chemoattractant's decay rate ($\varepsilon$), and secretion rate (s) by the cells for the contact inhibition model. The grey regions represent the 95% confidence intervals. None of the parameters, except for adhesion, affected the mean tip cell overtake rate per sprout significantly. As a rough estimate, all 95% confidence intervals overlap for the tip cell overtake rates. To quantitatively illustrate this, the mean tip cell overtake rate for T=0.5 compared to T=2 are not significantly different with a p-value of 0.901 for the contact inhibition model based on a Welch's t-test.

**Contact inhibition model**

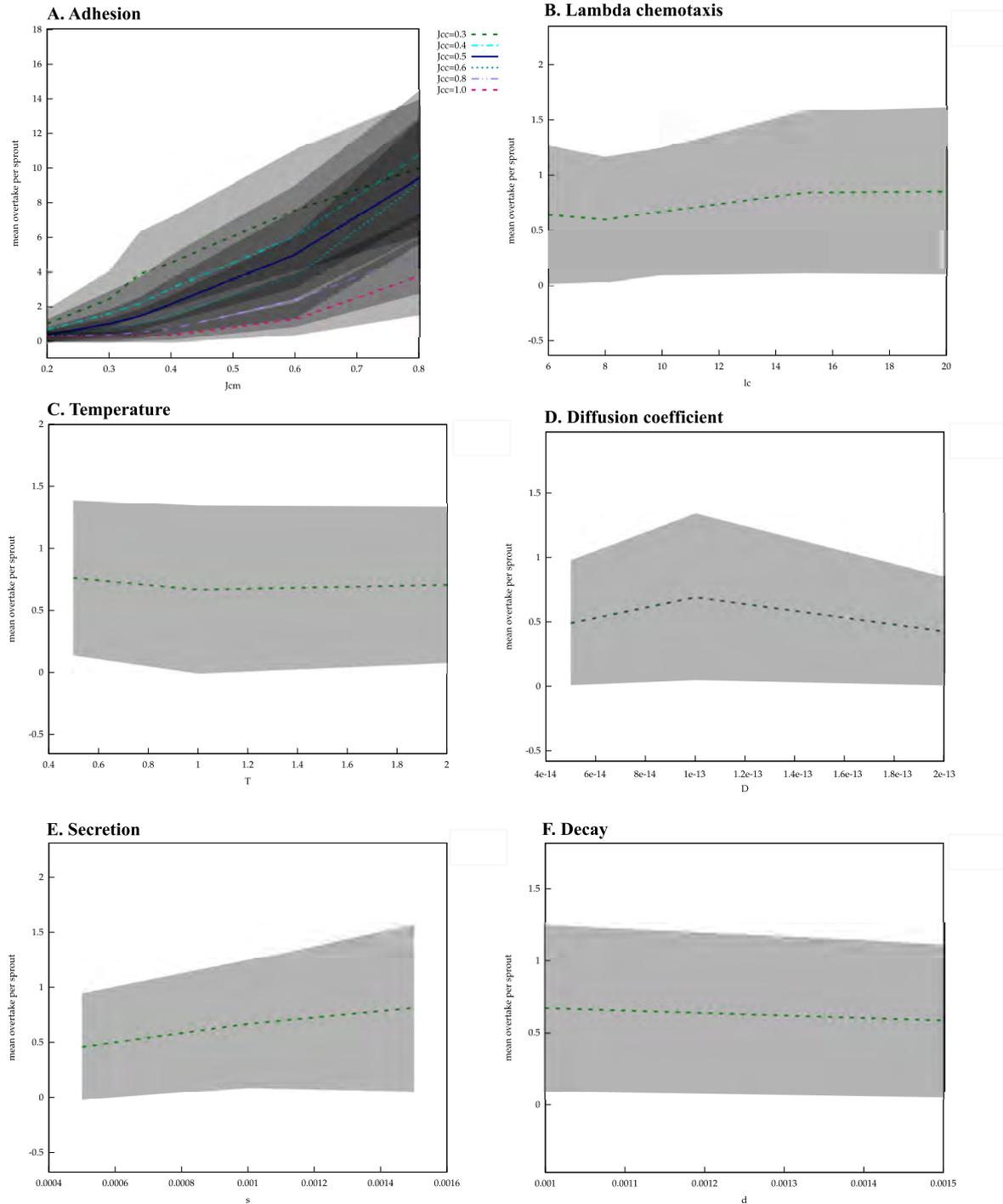

**Figure S2: Sensitivity of tip cell overtaking in cell elongation model [see next page]**
The mean overtake rate per sprout, based on 15 independent simulations, is plotted against cell-cell adhesion ($J_{cell,cell}$) and cell-ECM adhesion ($J_{cell,ECM}$), sensitivity to the auto-secreted chemoattractant ($\lambda_c$), the cellular temperature ($\mu$), the diffusion constant of the chemoattractant (D), the chemoattractant's decay rate (d), secretion rate (s), and the length of the cell (target length $L_l$ and cell elasticity $\lambda_l$) for the cell elongation model. The grey regions represent the 95% confidence intervals. None of the parameters affected the mean tip cell overtake rate per sprout significantly. As a rough estimate, all 95% confidence intervals overlap for the tip cell overtake rates. To quantitatively illustrate this, the mean tip cell overtake rate for T=0.5 compared to T=2 are not significantly different with a p-value of 0.093 for the cell elongation model based on a Welch's t-test.

# Cell elongation model

### A. Adhesion
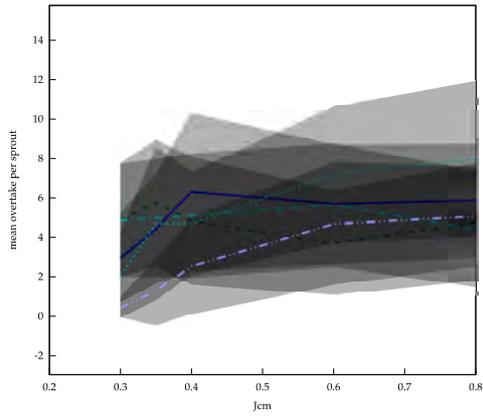

### B. Lambda chemotaxis
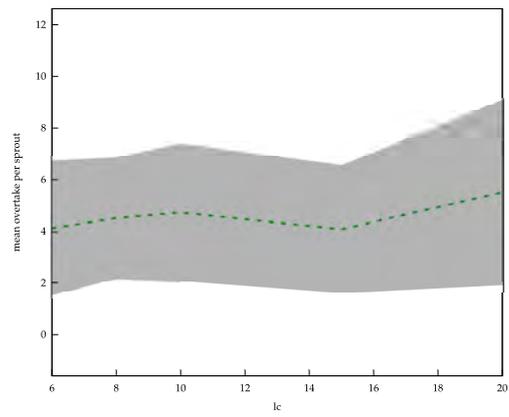

### C. Temperature
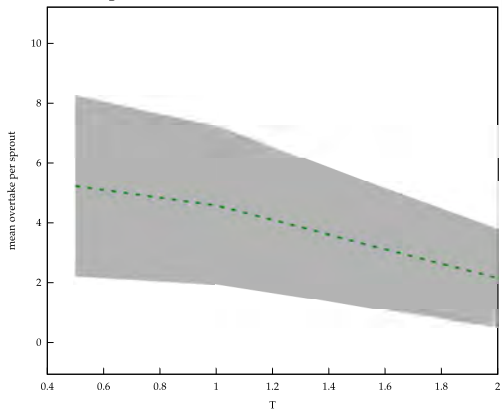

### D. Diffusion coefficient
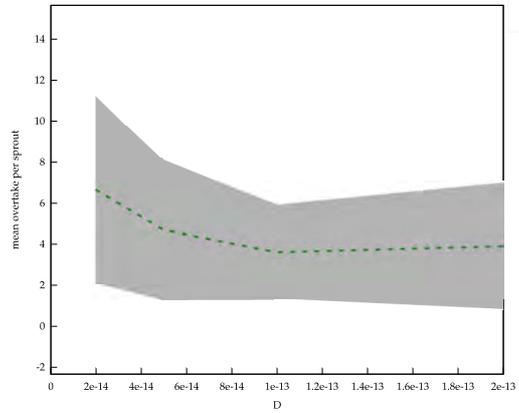

### E. Secretion
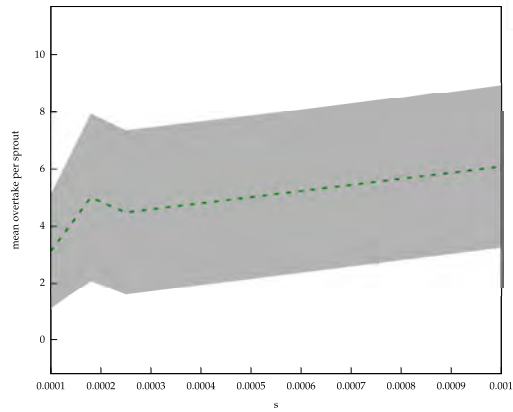

### F. Decay
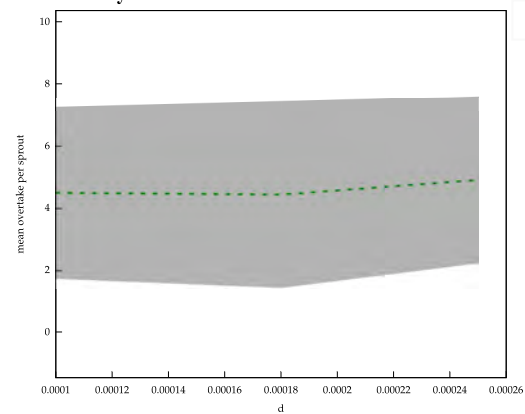

### G. Length
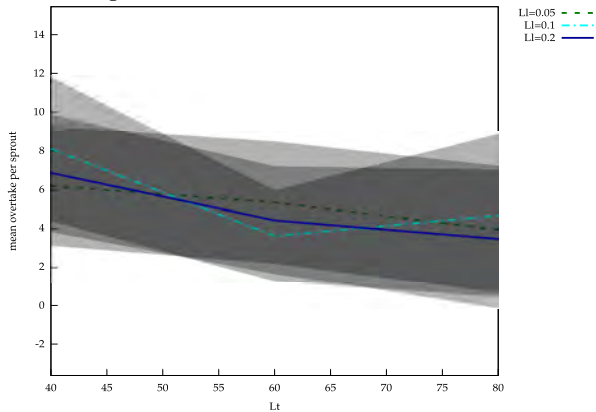

**Figure S3: Dll4 patterning by tip cell selection in the cell elongation model.**
Patterning of tip and stalk cells in a simulation of the cell elongation model. (A-C) Enlarged view of a sprout in which branching occurs in time, at the location of the white circle in panel A. (D-F) Enlarged view of two fusing sprouts (anastomosis) in time, indicated by the white circle in panel D. (G-I) Enlarged view of a sprout in which tip cell competition occurs in time at the location of the white circle in the panel G. The cell annotated with a square overtakes the tip cell position from the cell annotated with a star.

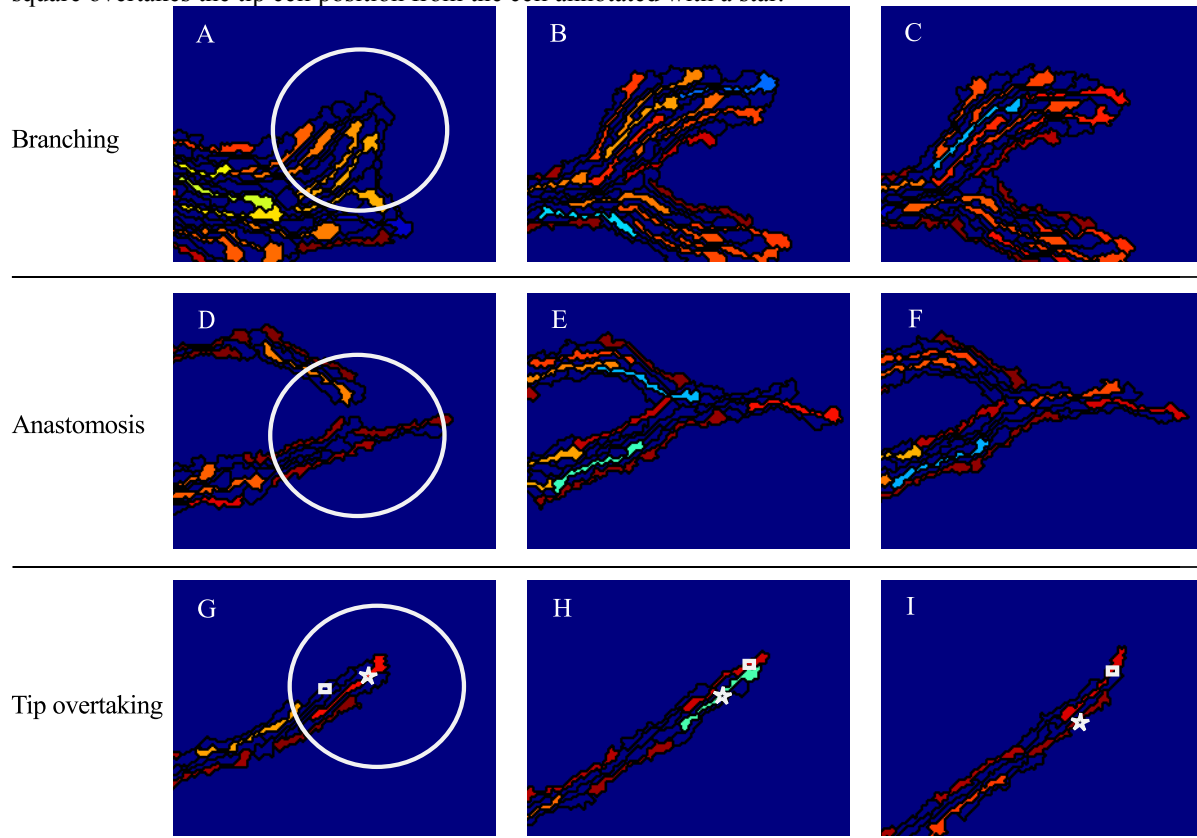

**Figure S4: Effect of VEGF gradients on the mean overtake rate per sprout.**
The mean overtake rate per sprout was calculated from ten simulations with the contact inhibition model in which only tip cells have a chemotactic sensitivity ($\lambda_{c,VEGF}=0.1$) to an external VEGF field. The different lines represent different shapes of the gradients of the external VEGF field ranging from concentration 0 to 1, which was uniformly spread over the grid, or increased from left to right over the grid in a linear, exponential or sigmoidal fashion. The mean overtake rate per sprout is plotted against the percentage of *Vegfr2* haploid cells in a mixed spheroid of WT cells and *Vegfr2* haploids. The grey regions represent the 95% confidence intervals. The mean overtake rate per sprout is not significantly different for distinct gradients of VEGF.

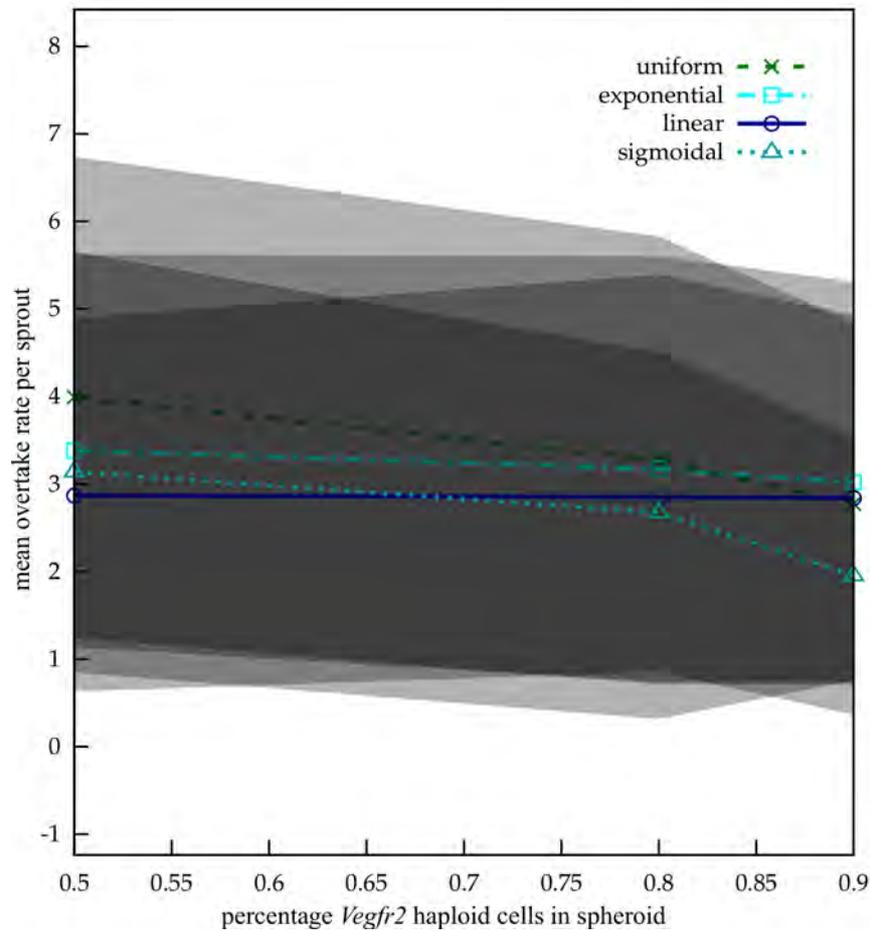

**Table S1: Effect of VEGF gradients on the sprout tip occupancy by WT cells.**
The mean (out of 10 simulations) occupancy of sprout tips by WT cells at the end of a simulation with the contact inhibition model with differential adhesion between tip and stalk cells, in which only tip cells have a chemotactic sensitivity ($\lambda_{c,VEGF}$=0.1) to an external VEGF field, is listed for different VEGF gradient shapes (columns) and for different ratios of WT and *Vegfr2* haploids in the spheroid (rows). The columns represent different shapes of the gradients of the external VEGF field ranging from concentration 0 to 1, which was uniformly spread over the grid, or increased from left to right over the grid in a linear, exponential or sigmoidal fashion. The p-values represent the probability that the total number of simulated sprouts were occupied by at least the indicated percentage of WT cells when assuming only random motion (calculated with a binomial distribution, with n the number of sprouts, k the number of sprouts occupied by WT cells, and p the mixing ratio.

**Table S2: Effect of VEGF gradient on cell trajectory data.**
Anterograde coordination, retrograde coordination, and the directional motility is listed for cells in the contact inhibition model (average of ten simulations) with differential adhesion between tip and stalk cells,, in which only tip cells have a chemotactic sensitivity ($\lambda_{c,VEGF}$=0.1) to an external VEGF field. The simulations were initialized with a mix of WT cells and *Vegfr2* haploids in a 1:1 ratio. The columns represent different shapes of gradients of the external VEGF field ranging from concentration 0 to 1, which was uniformly spread over the grid, or increased from left to right over the grid in a linear, exponential or sigmoidal fashion.

**Table S3: Parameter values of the contact inhibition model and the cell elongation model**

**Table S4: Parameter values VEGF-Dll4-Notch signaling model**
Dimensional units: decay rates, $\gamma_N$, $\gamma_D$, $\gamma_S$, $\gamma_R$, $\gamma_A$ are per 30 seconds (1 MCS = 30 s), production rates $\beta_N$, $\beta_D$, $\beta_{Dc}$, $\beta_R$, $\beta_{RC}$ in RU/(30 seconds) and affinities $k_S$, $k_D$, $k_A$ in RU · 30 seconds. Here Relative Units (RU) replace concentrations which are unknown.

**Table S1: Effect of VEGF gradients on the sprout tip occupancy by WT cells.**

| Ratio WT:*Vegfr2*[+/-] | Uniform | Linear | Sigmoidal | Exponential |
|---|---|---|---|---|
| **1:1** | 86 (p=1.1·10⁻⁷) | 71 (p=1.9·10⁻³) | 78 (p=7.8·10⁻⁵) | 75 (p=3.6·10⁻⁴) |
| **1:4** | 36 (p=8.7·10⁻³) | 57 (p=1.2·10⁻⁸) | 53 (p=4.6·10⁻⁷) | 49 (p=3.3·10⁻⁶) |
| **1:9** | 17 (p=9.3·10⁻²) | 13 (p=3.3·10⁻¹) | 19 (p=4.1·10⁻²) | 25 (p=1.2·10⁻³) |

**Table S2: Effect of VEGF gradient on cell trajectory data.**

|  | Uniform | Linear | Sigmoidal | Exponential |
|---|---|---|---|---|
| **Coordination anterograde** | 0.14 (±0.03) | 0.14 (±0.03) | 0.14 (±0.02) | 0.14 (±0.02) |
| **Coordination retrograde** | 0.14 (±0.03) | 0.14 (±0.03) | 0.14 (±0.02) | 0.14(±0.02) |
| **Directional motility** |  |  |  |  |
| - percentage anterograde | 48 | 29 | 35 | 22 |
| - percentage retrograde | 48 | 67 | 63 | 75 |
| - percentage stopped | 4 | 3 | 2 | 3 |

**Table S3: Parameter values of the contact inhibition model and the cell elongation model**

| Parameter | Description | Value in contact inhibition model | Value in cell elongation model | Unit |
|---|---|---|---|---|
| $\mu$ | Cellular temperature | 1 | 1 | - |
| $A$ | Target cell size | 50 | 100 | lattice sites |
| $\lambda_A$ | Cell elasticity | 0.5 | 1 | - |
| $J_{cell,ECM}$ | Cell-ECM adhesion | 0.4 | 0.35 | - |
| $J_{cell,cell}$ | Cell-cell adhesion | 0.8 | 0.5 | - |
| $\lambda_c$ | Sensitivity to the chemoattractant | 10 | 10 | - |
| $\alpha$ | Secretion rate | 1·10⁻³ | 1.8·10⁻⁴ | s⁻¹ |
| $\varepsilon$ | Decay rate | 1·10⁻³ | 1.8·10⁻⁴ | s⁻¹ |
| $D$ | Diffusion coefficient | 1·10⁻¹³ | 1·10⁻¹³ | m²/s |
| $H_{connectivity}$ | Connectivity | 1·10⁸ | 1·10⁸ | - |
| $\lambda_l$ | Cell length elasticity | - | 0.1 | - |
| $L$ | Target cell length | - | 60 | lattice sites |

Table S4: Parameter values VEGF-Dll4-Notch signaling model

| Parameter | Description | Value |
|---|---|---|
| $\beta_N$ | Production rate Notch | 1 |
| $\gamma_N$ | Decay rate Notch | 0.1 |
| $\beta_D$ | Variable production rate Dll4 | 5 |
| $\gamma_D$ | Decay rate Dll4 | 0.1 |
| $\beta_{Dc}$ | Constitutive production rate Dll4 | 0.1 |
| $\gamma_S$ | Decay rate NICD | 0.1 |
| $\alpha_S$ | Production rate NICD | 100 |
| $k_S$ | Hill constant that relates Dll4-Notch signaling to NICD production | 3000 |
| $n_S$ | Hill constant that relates Dll4-Notch signaling to NICD production | 2 |
| $m_D$ | Hill constant that relates NICD to Dll4 production | 1 |
| $k_t$ | Trans-signaling coefficient | 80 |
| $k_c$ | Cis-signaling coefficient | 10 |
| $d$ | Scaling constant | 2 |
| $k_D$ | Hill constant that relates VEGF signaling activity to Dll4 production | 130000 |
| $\alpha_D$ | Production rate of Dll4 depending on VEGF signaling activity | 15 |
| $\beta_R$ | Variable VEGFR2 production rate | 2 |
| $\beta_{RC}$ | Constant VEGFR2 production rate | 0.01 |
| $\gamma_R$ | VEGFR2 decay rate | 0.3 |
| $m_R$ | VEGF signaling activity | 2 |
| $n_D$ | Hill constant that relates VEGF signaling activity to Dll4 production | 2 |
| $\alpha_A$ | Production rate of VEGF signaling activity | 100 |
| $n_A$ | Hill constant that relates VEGF-VEGFR2 binding to VEGF signaling activity | 2 |
| $k_A$ | Hill constant that relates VEGF-VEGFR2 binding to VEGF signaling activity | 30 |
| $\gamma_A$ | Decay rate of VEGF signaling activity | 0.1 |